\documentclass[twoside,english]{iopart}
\usepackage[T1]{fontenc}
\usepackage[latin9]{inputenc}
\usepackage{geometry}
\usepackage{booktabs}
\usepackage{graphicx}

\makeatletter

\providecommand{\tabularnewline}{\\}

\usepackage{iopams}
\usepackage{setstack}

\usepackage[numbers, sort&compress]{natbib}

\newcommand{\eqref}[1]{(\ref{#1})}

\usepackage[singlelinecheck=on,justification=justified]{caption}
\usepackage[load-configurations=abbreviations,range-phrase = ~-~,range-units = single,separate-uncertainty = true]{siunitx}
\usepackage[siunitx]{circuitikz}
\RequirePackage{subscript}
\usepackage{ifpdf}
\ifpdf
\usepackage{epstopdf}
\fi
\usepackage[labelfont=bf]{caption}
\newcommand{\YBCO}{\emph{RE}BCO~}

\newcommand{\sym}[2]{\ensuremath{#1_\textrm{#2}}}

\newcommand{\n}[1]{\textrm{#1}}

\usepackage{hyperref}

\makeatother

\usepackage{babel}
\begin{document}

\title[Temperature- and Field Dependent Characterization of a CORC Cable]{Temperature- and Field Dependent Characterization of a Conductor
on Round Core Cable.}

\author{{\Large{}C Barth{\Large{}\textsuperscript{1,2}}, D C van der Laan{\Large{}\textsuperscript{3,4}},
N Bagrets{\Large{}\textsuperscript{1}}, C M Bayer{\Large{}\textsuperscript{1}},
K-P Weiss{\Large{}\textsuperscript{1}} and C Lange{\Large{}\textsuperscript{1}} }}

\address{{\large{}{\large{}\textsuperscript{1}} Institute for Technical Physics
(ITEP), Karlsruhe Institute of Technology (KIT), Germany}}

\address{{\large{}{\large{}\textsuperscript{2}} Department of Condensed Matter
Physics (DPMC), University of Geneva, Switzerland}}

\address{{\large{}{\large{}\textsuperscript{3}} Advanced Conductor Technologies
LLC, 3082 Sterling Circle, Boulder, USA}}

\address{{\large{}{\large{}\textsuperscript{4}} Department of Physics, University
of Colorado, Boulder, USA}}

\ead{{\large{}christian.barth@unige.ch}}
\begin{abstract}
The Conductor on Round Core (CORC) cable is one of the major high
temperature superconductor cable concepts combining scalability, flexibility,
mechanical strength, ease of fabrication and high current density;
making it a possible candidate as conductor for large, high field
magnets. To simulate the boundary conditions of such magnets as well
as the temperature dependence of Conductor on Round Core cables a
\SI{1.16}{\metre} long sample consisting of 15, \SI{4}{\milli\metre}
wide SuperPower \YBCO tapes was characterized using the ``FBI''
(force - field - current) superconductor test facility of the Institute
for Technical Physics (ITEP) of the Karlsruhe Institute of Technology
(KIT). In a five step investigation, the CORC cable's performance
was determined at different transverse mechanical loads, magnetic
background fields and temperatures as well as its response to swift
current changes. In the first step, the sample's \SI{77}{\kelvin},
self-field current was measured in a liquid nitrogen bath. In the
second step, the temperature dependence was measured at self-field
condition and compared with extrapolated single tape data. In the
third step, the magnetic background field was repeatedly cycled while
measuring the current carrying capabilities to determine the impact
of transverse Lorentz forces on the CORC cable sample's performance.
In the fourth step, the sample's current carrying capabilities were
measured at different background fields (\SIrange{2}{12}{\tesla})
and surface temperatures (\SIrange{4.2}{51.5}{\kelvin}). Through
finite element method (FEM) simulations, the surface temperatures
are converted into average sample temperatures and the gained field-
and temperature dependence is compared with extrapolated single tape
data. In the fifth step, the response of the CORC cable sample to
rapid current changes (\SI{8.3}{\kilo\ampere\per\second}) was observed
with a fast data acquisition system. During these tests, the sample
performance remains constant, no degradation is observed. The sample's
measured current carrying capabilities correlate to those of single
tapes assuming field- and temperature dependence as published by the
manufacturer. 
\end{abstract}


\submitto{\SUST }

\maketitle

\section{Introduction}

Second generation high temperature superconductors (HTS) are the rare-earth-barium-copper-oxide
(\emph{RE}BCO) tapes, often referred to as coated conductors. They
are of thin tape shape, commonly with widths of \SIrange{3}{15}{\milli\metre}
and thicknesses in the \SIrange{50}{200}{\micro\metre} range. The
mechanical properties as well as the performance in strong magnetic
background fields surpasses first generation high temperature superconductors
making \YBCO tapes a desired conductor for rotating machinery, fusion
magnets and high field magnets.

\subsection{High temperature superconductor cable concepts}

Due to their tape geometry, the cabling methods established for low
temperature superconductors are not applicable. Different approaches
are necessary. There are at present five experimentally proven concepts
of how to combine several \YBCO tapes into flexible, mechanically
strong cables able to carry \si{\kilo\ampere} currents in strong
background fields. 
\begin{description}
\item [{Roebel~Assembled~Coated~Conductor~(RACC)~cables}] are being
developed at the Karlsruhe Institute of Technology (KIT), Germany,
and Industrial Research Limited (IRL), New Zealand \cite{Goldacker2006,Long2011}. 
\item [{Coated~Conductor~Rutherford~Cables~(CCRC)}] are being developed
at the Karlsruhe Institute of Technology (KIT), Germany \cite{Schlachter2011,Kario2013}.
\item [{Conductor~on~Round~Core~(CORC)~cables}] are being developed
at Advanced Conductor Technologies, USA, and the University of Colorado,
USA \cite{VanderLaan2009,VanderLaan2011a,VanderLaan2012a,VanderLaan2013}. 
\item [{Twisted~Stacked-Tape~Cables~(TSTC)}] are being developed at
the Massachusetts Institute of Technology, USA \cite{Takayasu2009,takayasu2011,Takayasu2012,Takayasu2012a,Chiesa2014}
and at the Italian National Agency for New Technologies, Energy and
Sustainable Economic Development (ENEA), Italy \cite{Celentano2014}.
\item [{Round~Strands~Composed~of~Coated~Conductor~Tapes~(RSCCCT)}] are
being developed at the Centre de Recherches en Physique des Plasmas
(CRPP) of the \'Ecole Polytechnique F\'ed\'erale de Lausanne (EPFL)
\cite{Uglietti2013}. 
\end{description}
There are significant differences in the arrangement of the tapes,
the tape consumption, the transposition, the mechanical properties
as well as the in-field performance of the HTS cable concepts. Their
applicability is thus depending on boundary conditions.

\subsection{Conductor on Round Core Cables\label{sec:CORC}}

\begin{figure*}[tbh]
\begin{centering}
\includegraphics[width=0.75\textwidth]{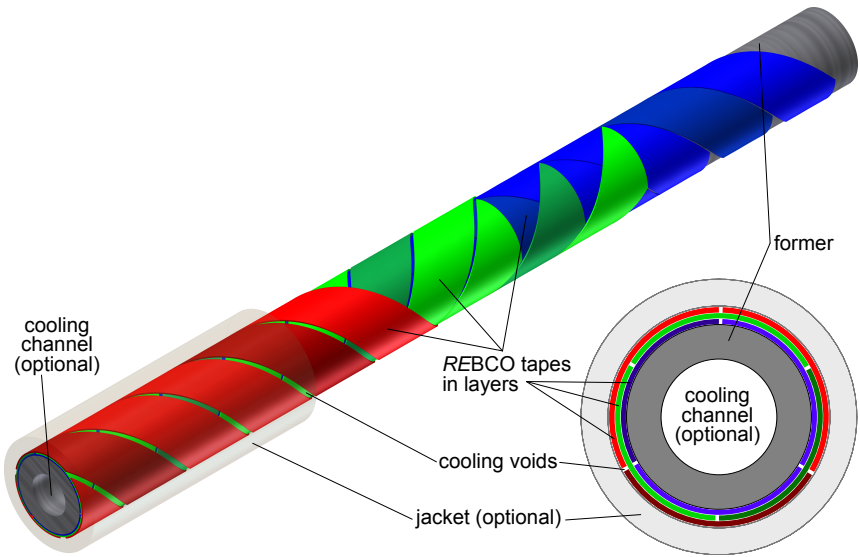}
\par\end{centering}

\caption{Schematic drawing of a CORC cable. \YBCO tapes are tightly wound
around a round former with a small diameter. The winding direction
between the layers is reversed. The cable can optionally be fitted
with a jacket of structural material for increased mechanical stabilization
and with a hollow former for forced flow cooling. Picture from~\cite{Barth-PhD}.\label{fig:CORC}}
\end{figure*}

Conductor on Round Core cables consist of \YBCO tapes which are tightly
wound around a round former in several layers using a winding angle
between \ang{30} and \ang{60}. The winding directions between each
layer are reversed. In contrast to HTS power cables, formers with
significantly smaller diameters are used. Existing CORC cables either
utilize a copper rod or a copper power cable of \SI{5.5}{\milli\meter}
(or less) diameter as a former \cite{VanderLaan2011a}. In this geometry
there are only three \SI{4}{\milli\metre} wide tapes per layer in
the three inner layers. In subsequent layers, the number of tapes
increases due to the increase in winding diameter. The former provides
electrical and mechanical stabilization. For additional mechanical
stabilization the cable can be fitted with a jacket of structural
material. Hollow formers, allowing forced flow cooling of the cable,
are also possible. This cable concept is shown in schematic drawing
in figure~\ref{fig:CORC}. 

In CORC cables, the \YBCO tapes are wound, with the superconducting
layer in compression, to small diameters. This is possible without
major reduction of their current carrying capabilities due to winding
angles close to \SI{45}{\degree}, symmetrically distributing the
winding strains onto the a-axis (100 direction) and b-axis (010 direction)
of all \YBCO crystals' unit cells. This arrangement suppresses the
impact of the winding strain on the current carrying capabilities,
the so called reversible strain effect of the superconductor tapes
\cite{VanderLaan2011-YBCO-anisotropy}. However, even though the tapes
are wound around the former in different winding directions, the transposition
of the \YBCO tapes in the CORC cable concept is only partial. This
makes the response of CORC cables to swiftly changing currents highly
interesting. The periodically changing orientation (due to the winding)
of the \YBCO tapes makes the cable's current depending on the field
component with the strongest critical current degradation. This however
has two benefits in applications. Firstly, the field dependence of
CORC cables is isotropic, the position in e.g. a solenoid magnet is
of no importance. Secondly, the engineering current density \sym{J}{e}
of CORC cables is expected to increase rapidly with enhanced pinning,
likely without an increase in conductor cost. In the following, a
CORC cable sample is experimentally characterized in detail.

\section{Experimental}

\subsection{Sample parameters\label{sec:Sample-parameters}}

The investigated CORC cable sample was produced by Advanced Conductor
Technologies. It was of \SI{1.16}{\metre} length and consisted of
15, \SI{4}{\milli\metre} wide SuperPower \YBCO tapes stabilized
with \SI{20}{\micro\metre} thick layers of copper (Cu) plating on
both sides (SCS4050). The tapes were arranged in five layers (three
tapes per layer) and were wound around a round former of \SI{5.5}{\milli\metre}
diameter with a winding angle of ca. \ang{45}. The former was a conventional
stranded power cable which has been soldered into the terminations
of the CORC cable sample. It provided mechanical and electrical stabilization.
Three pairs of voltage taps were soldered to the \YBCO tapes of the
the outer layer and one pair was attached to the copper terminations
of the sample. All sample parameters are summarized in table~\ref{tab:Sample-parameters}. 

\begin{table*}[tbph]
\caption{Parameters of the investigated CORC cable sample.\label{tab:Sample-parameters}}

\centering{}%
\begin{tabular}{cc}
\toprule 
{\footnotesize{}parameter} & {\footnotesize{}CORC cable sample}\tabularnewline
\midrule
\midrule 
{\footnotesize{}sample length} & {\footnotesize{}\SI{1.16}{\metre} including terminations, \SI{0.96}{\metre}
between terminations}\tabularnewline
\midrule 
{\footnotesize{}superconductor} & {\footnotesize{}\SI{4}{\milli\metre} wide, copper stabilized from
SuperPower (SCS4050) - no advanced pinning}\tabularnewline
\midrule 
{\footnotesize{}average tape \sym{I}{c}} & {\footnotesize{}\SI{127.57}{\ampere} at \SI{77}{\kelvin}, s.f. and
\SI{161.42}{\ampere} at \SI{4.2}{\kelvin}, \SI{12}{\tesla}}\tabularnewline
\midrule 
{\footnotesize{}number of tapes} & {\footnotesize{}15 tapes in 5 layers (3 tapes per layer)}\tabularnewline
\midrule 
{\footnotesize{}twist pitch $\tau$} & {\footnotesize{}\SI{17}{\milli\metre}}\tabularnewline
\midrule 
{\footnotesize{}termination} & {\footnotesize{}all tapes individually soldered to cone shaped copper
contacts}\tabularnewline
\midrule 
{\footnotesize{}mechanical stabilization} & {\footnotesize{}stranded insulted former (power cable) with \SI{5.5}{\milli\metre}
diameter}\tabularnewline
\midrule 
{\footnotesize{}electrical stabilization} & {\footnotesize{}{\footnotesize{}\SI{20}{\micro\metre}} thick Cu stabilization
of the tapes + Cu in the strands of former}\tabularnewline
\midrule 
{\footnotesize{}voltage tap pairs} & {\footnotesize{}3 pairs at tapes in the outer layer + 1 pair at the
copper contacts (the terminations)}\tabularnewline
\bottomrule
\end{tabular}
\end{table*}

The average critical current of the used \YBCO tapes was determined
after the field- and temperature dependent experiments by characterizing
tapes from the same batch at \SI{77}{\kelvin}, self-field and at
\SI{4.2}{\kelvin}, \SI{12}{\tesla}. The field was applied perpendicular
to the tape surface.

\subsection{Measurement procedure}

In a five step approach, the CORC cable sample's performance was investigated
at different transverse mechanical loads, magnetic background fields
and temperatures.
\begin{itemize}
\item Critical current measurements at \SI{77}{\kelvin}, self-field in
a liquid nitrogen bath (subsection~\ref{sec:77K-characterization}).
\item Temperature variable measurements at self-field conditions utilizing
the variable temperature insert of the FBI (force - field - current)
test facility (subsection~\ref{sec:T-variable}).
\item Impact of transverse mechanical loads generated by Lorentz forces
in magnetic fields cycles of \SIrange{2}{12}{\tesla}. Degradations
due to the increasing Lorentz forces will be made visible in these
measurements by comparing the obtained critical currents of different
load cycles (subsection~\ref{sec:Load-cycling}).
\item Critical current measurements at different fields and temperatures
to determine the magnetic field and temperature dependence of CORC
cables. The gained data is compared with single tape data published
by the superconductor's manufacturer (subsection~\ref{sec:T-variable+field-variable}).
\item Response of the CORC cable sample to rapidly changing currents by
measuring the voltage drop over the whole cable at a current ramp
rate of \SI{8.3}{\kilo\ampere\per\second} (subsection~\ref{sub:CORC-high-ramp-rates}). 
\end{itemize}

\subsection{Test setup\label{sec:Test-setup}}

\begin{figure*}[tbh]
\begin{centering}
\includegraphics[width=1\textwidth]{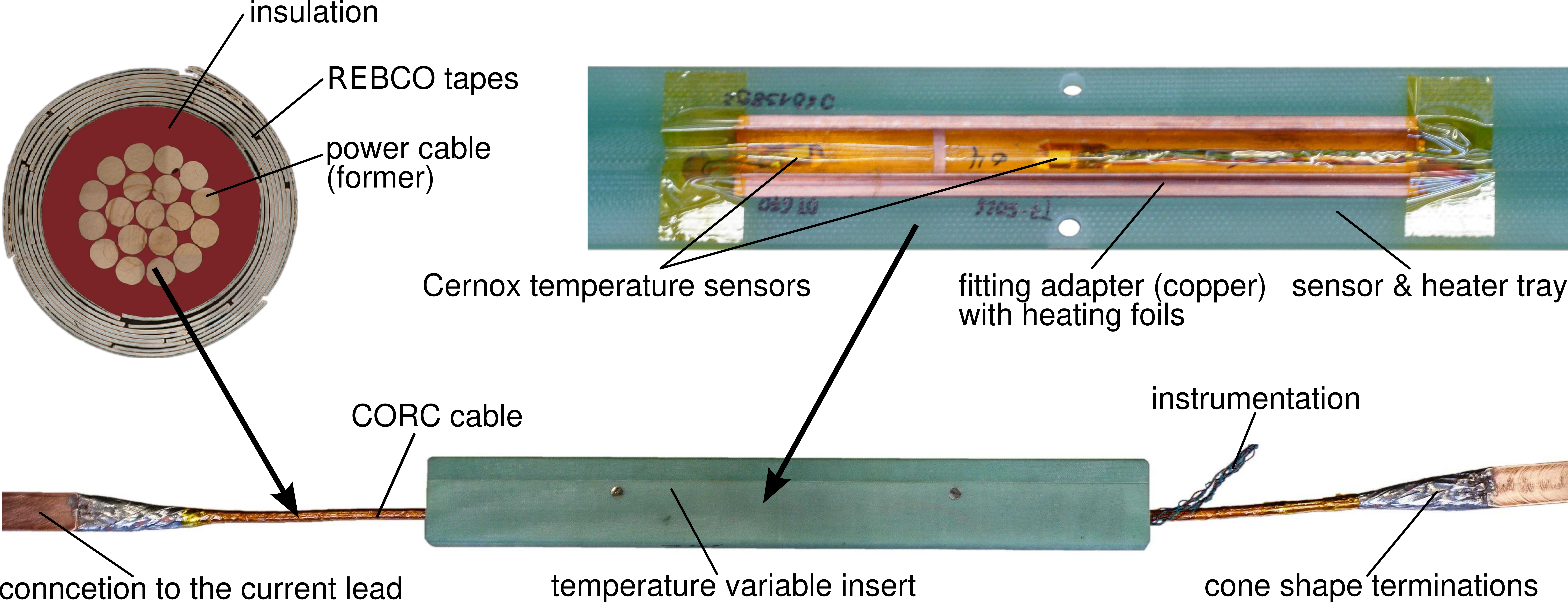}
\par\end{centering}

\caption{CORC cable sample in the variable temperature insert used in the magnetic
field- and temperature dependent measurements. Picture from~\cite{VanderLaan2011a,Barth-PhD}.\label{fig:CORC-sample}}
\end{figure*}

The field- and temperature dependent characterizations were performed
using the FBI superconductor test facility of the Institute for Technical
Physics (ITEP) of the Karlsruhe Institute of Technology (KIT) \cite{Bayer2014}.
The test facility consisted of:
\begin{description}
\item [{Current~source:}] A low noise DC power supply provided up to \SI{10}{\kilo\ampere}
of current to the sample. The measurements were performed at quasi-constant
current; the current was increased slowly with a ramp rate of \SI{20}{\ampere\per\second}
from step to step. On each of the steps, the current was kept constant
while the voltages were measured using low noise nano-voltmeters. 
\item [{Magnet:}] A superconducting split coil magnet delivered magnetic
background fields up to \SI{12}{\tesla}. The field was orientated
perpendicular to the sample axis and the current. Approximately 4
transposition lengths of the sample were inside the \SI{70}{\milli\metre}
long high field region (field deviation less than \SI{3}{\percent}
from the peak value). Above and below the magnet, there were current
leads connecting the sample to the power supply.
\item [{Variable~temperature~insert:}] In the variable temperature insert,
the central \SI{100}{\milli\metre} of the sample were heated with
heating foils while being thermally insulated from the helium bath
with glass-fiber-reinforce-plastics (GFRP) G10 ``cryostat''. The
heated section of the sample and the high field region of the magnet
were aligned. The heating system fitted tightly around the sample
and was sealed with bees wax and adhesive tape at the upper and lower
end to minimizes helium boil off. This system has been shown to exhibit
fast response (less than \SI{1}{\minute}) and stable temperature,
however there was a temperature gradient. The temperature gradient,
as demonstrated on a CICC dummy, was low in longitudinal direction
and was pronounced in radial direction \cite{Barth2014-TSTC-temp}.
As it was not possible to place temperature sensors on the CORC cable's
inside, its temperature distribution was therefore simulated (subsection~\ref{sec:T-distribution}).
There were two Cernox temperature sensors in direct contact with the
sample's surface, their average was used as input for the model and
is referred to as the surface temperature \sym{T}{surface}. The CORC
cable sample equipped with the variable temperature insert is shown
in figure~\ref{fig:CORC-sample}. 
\end{description}

\subsection{Temperature distribution of the Conductor on Round Core cable\label{sec:T-distribution}}

\begin{figure}[tbh]
\begin{centering}
\includegraphics[width=0.5\textwidth]{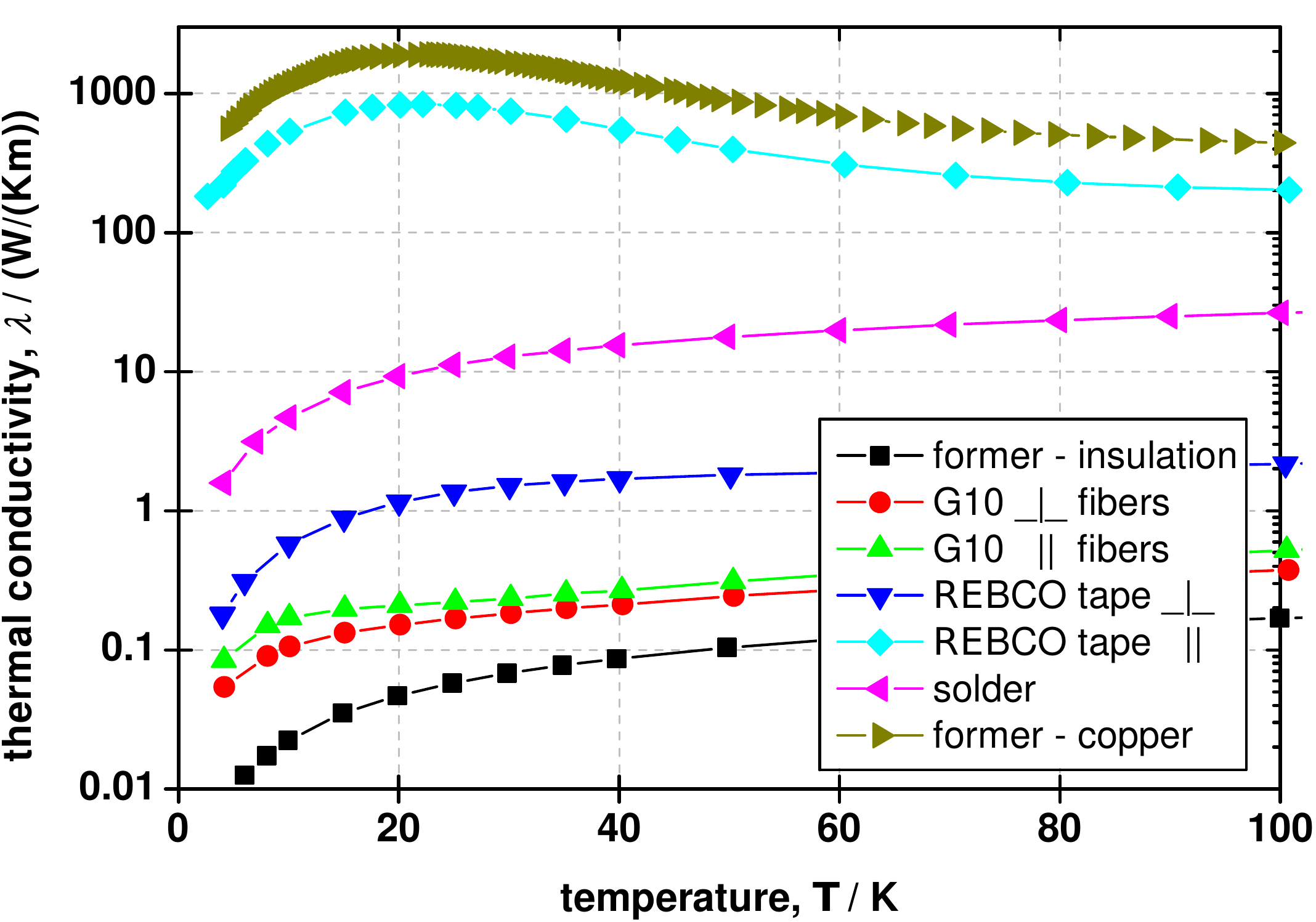}
\par\end{centering}

\caption{Thermal conductivity of the CORC cable sample's constituent materials.
Data from \cite{Bagrets2014,Bagrets2014a,Barth-PhD}.\label{fig:TC}}
\end{figure}

The heated section of the sample was modeled in full scale in 3D using
the direction- and temperature dependent thermal conductivities of
all constituent materials as shown in figure~\ref{fig:TC} with the
commercial software package ``Comsol'' \cite{Comsol-SUST}. 

\begin{figure*}[tbph]
\begin{centering}
\includegraphics[width=1\textwidth]{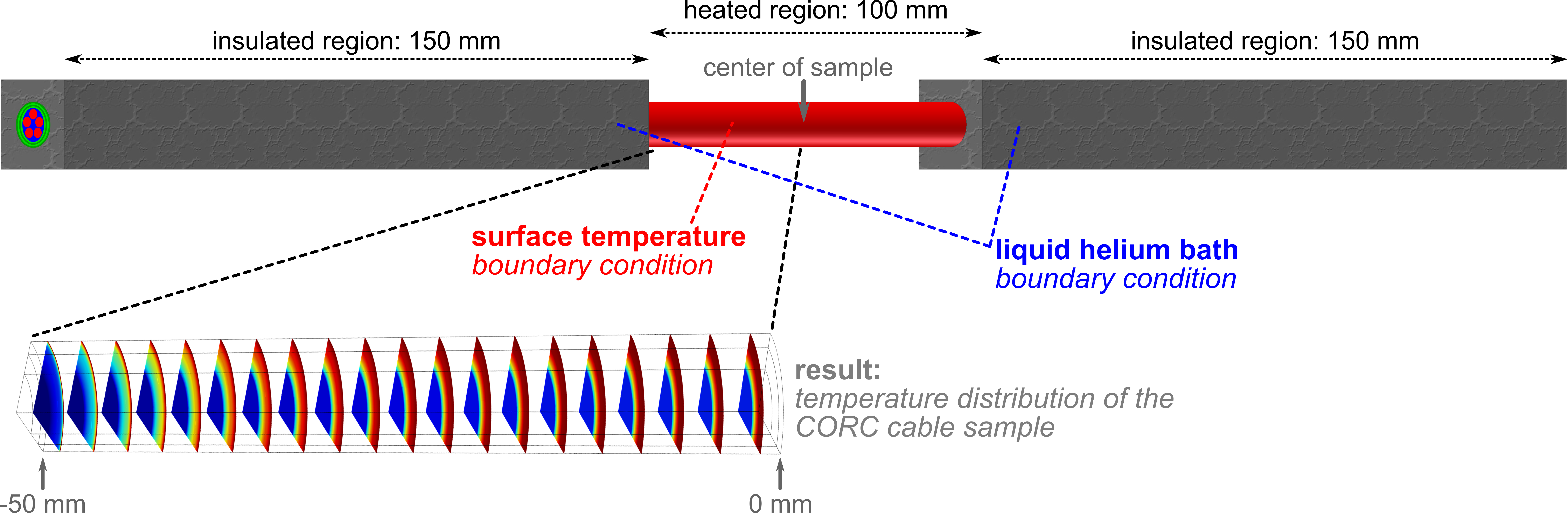}
\par\end{centering}

\caption{Schematic drawing of the FEM models used in the simulations of the
temperature distribution of HTS cables in the variable temperature
insert. An exemplary temperature distribution is shown as a color
gradient from cold (blue) to hot (red). The figure is not up to scale.\label{fig:FEM-t-distribution-method}}
\end{figure*}

In the model, temperature sources (the average of the two Cernox sensors:
surface temperature \sym{T}{surface}) and temperature sinks (the
helium bath: \SI{4.2}{\kelvin}) were imposed as boundary conditions.
For optimal resolution, different mesh configurations were used. The
\YBCO tapes were netted with a mapped mesh allowing the placement
of several nodes within the width of each tape's superconducting layer.
A free tetradic mesh, with much lower node density, was used in the
variable temperature insert. This simulation method is shown schematically
in figure~\ref{fig:FEM-t-distribution-method}.

\begin{figure}[tbh]
\begin{centering}
\includegraphics[width=0.5\textwidth]{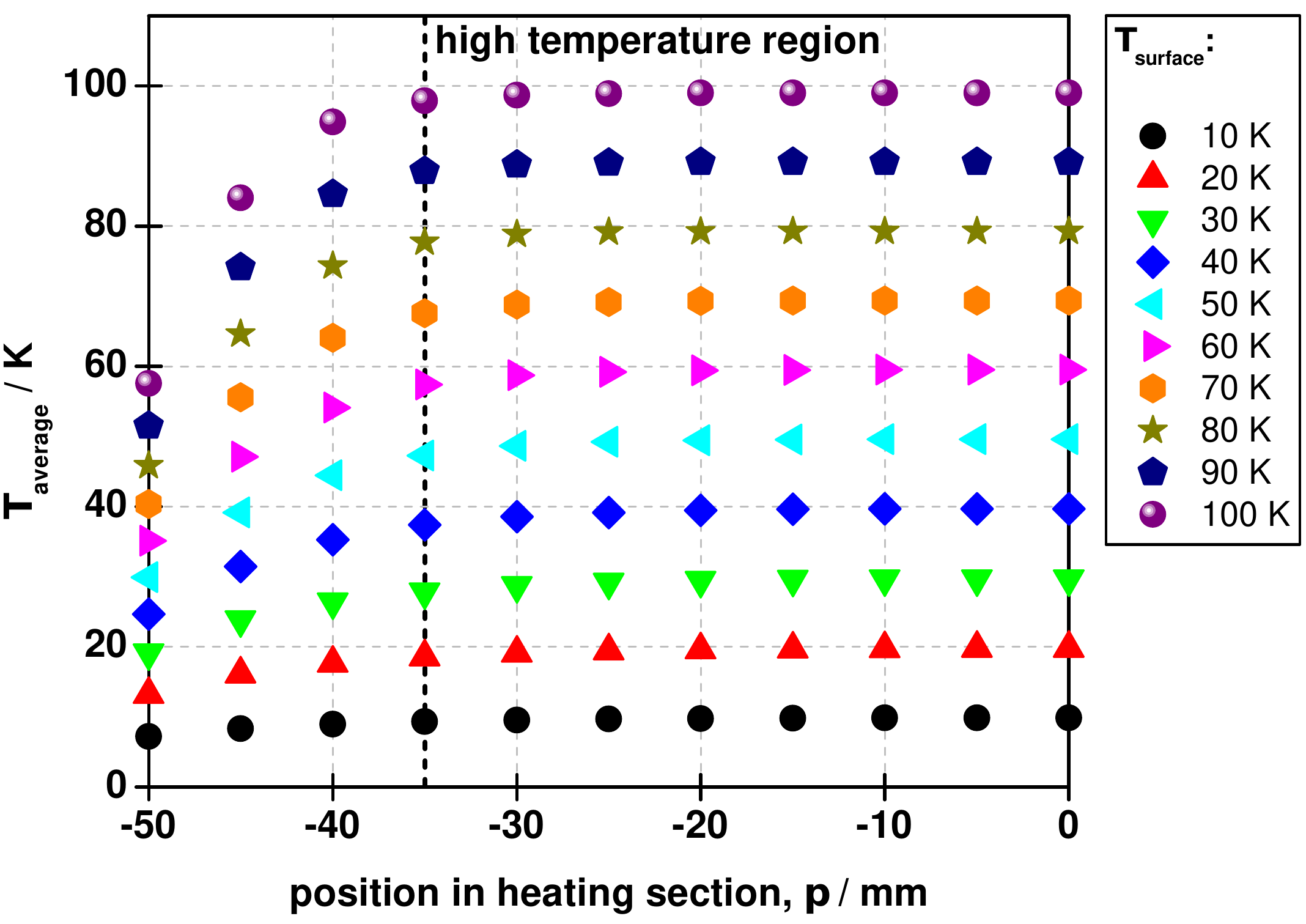}
\par\end{centering}

\caption{Simulated temperature distribution of the CORC cable sample. The average
temperature of the cable \sym{T}{average} is shown at different positions
in the heating section from the border (position: \SI{-50}{\milli\metre})
to the center (position: \SI{0}{\milli\metre}) of the \SI{100}{\milli\metre}
long heating section)\label{fig:CORC-T-distribution}}
\end{figure}

With this method the temperature distribution was calculated for different
surface temperatures \sym{T}{surface}. In regularly spaced positions
along the sample (each \SI{5}{\milli\metre}), the temperatures of
all tapes were averaged. In figure~\ref{fig:CORC-T-distribution},
this averaged cable temperature \sym{T}{average} is shown from the
center (position: \SI{0}{\milli\metre}) to the border of the heating
section (position: \SI{-50}{\milli\metre}) for surface temperatures
from \SI{10}{\kelvin} to \SI{100}{\kelvin}.

From \SI{-35}{\milli\metre} to the sample's center (position: \SI{0}{\milli\metre}),
the average temperature \sym{T}{average} was stable. Due to the system's
symmetry this corresponded to the CORC sample's central \SI{70}{\milli\metre}.
There was a pronounced drop of the average temperature further away
from the center (positions: \SI{<-35}{\milli\metre} and \SI{>35}{\milli\metre}).
The central \SI{70}{\milli\metre} were therefore the variable temperature
insert's active zone and are referred to in the following as the sample's
high temperature region corresponding to the magnet's high field region.
Within the temperature range of interest (surface temperature: \SI{<40}{\kelvin}),
the deviations from the average temperature in this zone were less
than \SI{\pm3}{\percent}. This distance of \SI{70}{\milli\metre}
was therefore a good assumption to define the working area of the
variable temperature insert as well as the magnet. In this zone, magnetic
field and temperature were assumed to be maximal as well as constant
and the \SI{70}{\milli\metre} were used to convert the measured voltages
into electric fields regardless of the voltage taps' separation. Outside
these \SI{70}{\milli\metre}, the sample was at lower fields and/or
temperatures. It was therefore away from the superconducting transition
without any contribution to the measured sample voltages.

\section{Results\label{sec:Results}}

\subsection{LN\textsubscript{2} characterization\label{sec:77K-characterization}}

In the first step of the CORC cable characterization, the current
carrying capabilities of the sample were measured in a liquid nitrogen
(LN\textsubscript{2}) bath at \SI{77}{\kelvin} in self-field conditions.
Due to the steep transitions and homogeneous contact resistances,
a \SI{1}{\micro\volt\per\centi\metre} criteria was used to determinate
of the critical currents in all following investigations. As there
was neither magnetic field nor heating, the actual distance between
the voltage taps was used to convert voltages into electric fields.
The critical currents gained from all available voltage taps are shown
in table~\ref{tab:CORC-77K-self-field}, subtracting any linear ohmic
contributions.

\begin{table}[tbh]
\caption{\SI{77}{\kelvin} (LN\textsubscript{2} bath), self-field current
carrying of the CORC cable sample for different voltage taps (\SI{1}{\micro\volt\per\centi\metre}
criteria). Linear ohmic contributions are subtracted.\label{tab:CORC-77K-self-field}}

\centering{}%
\begin{tabular}{ccccc}
\toprule 
{\footnotesize{}voltage tap:} & {\footnotesize{}tap1} & {\footnotesize{}tap2} & {\footnotesize{}tap3} & {\footnotesize{}terminations}\tabularnewline
\midrule
\midrule 
\textbf{\footnotesize{}\sym{I}{c}:} & {\footnotesize{}\SI{1752}{\ampere}} & {\footnotesize{}\SI{1652}{\ampere}} & {\footnotesize{}\SI{1757}{\ampere}} & {\footnotesize{}\SI{1638}{\ampere}}\tabularnewline
\bottomrule
\end{tabular}
\end{table}

Voltage taps are only available at the outer three of the 15 \YBCO
tapes of the sample. Even these tapes transition at slightly different
currents (\SIrange{1652}{1757}{\ampere}). Thus, the voltage drop
across the copper terminations (the sample's contacts) is most relevant
as it averages over the behavior of all tapes, it is therefore used
for the determination of the samples current carrying capabilities
in all following experiments. Using the copper contacts' voltage taps,
the \SI{77}{\kelvin}, self-field the critical current of sample is
\ensuremath{\sym{I}{c}\left(\SI{77}{\kelvin}, \n{self-field}\right)=\SI{1638}{\ampere}}
(see~\cite[p. 7]{EFDA-2012-06}). The used superconductor tapes have
an average \SI{77}{\kelvin}, self-field current of \SI{127.57}{\ampere}
(measured in LN\textsubscript{2} bath on several tapes of the same
\YBCO tape batch), resulting in a design current ($15\times$ \SI{127.57}{\ampere})
of the CORC cable of \SI{1913}{\ampere}. The sample's self-field
degradation is \SI{14.38}{\percent}.

\subsection{Temperature dependence\label{sec:T-variable}}

In the second step, the temperature dependence of the current carrying
capabilities of the CORC cable sample was investigated at self-field
conditions using the variable temperature insert. Critical currents
were determined with a \SI{1}{\micro\volt\per\centi\metre} criteria
using the voltage taps at the copper contacts. A distance of \SI{70}{\milli\metre}
(the variable temperature insert's high temperature region) was used
to convert voltages into electric fields. Outside this region, the
temperatures dropped quickly therefore avoiding any influence on the
measured voltages. The current bypass through the copper in the former
(the power cable) was determined to be negligible. In the sample's
former, there were $19$ copper strands of \SI{0.75}{\milli\metre}
diameter resulting in a total copper cross sectional area of \SI{8.32E-9}{\metre\squared}.
With a residual resistivity ratio (RRR) of $100$ and a specific resistivity
of copper at room temperature of \SI{16.78}{\nano\ohm\metre}, the
total electrical resistivity between the terminations (\SI{0.96}{\metre}
separation) was \SI{19}{\micro\ohm} at operating conditions. This
resulted in a current bypass of \SI{3.6}{\ampere} at ten times the
critical electric field (voltage of \SI{7E-5}{\volt} over the whole
sample). This calculation does not account for the contact resistance
between the terminations and the former and as the former was clamped
into the terminations and fixed with a stainless steel screw, thus
this in an upper limit of the real current bypass. The simulated temperature
distribution of the sample (figure~\ref{fig:CORC-T-distribution})
was averaged in the high temperature region leading to the the average
cable temperatures \sym{T}{average} for the experiment's measured
surface temperatures \sym{T}{surface}. These temperatures as well
as the statistic (derived from the averaging) and the systematic (from
the temperature measurement) uncertainties are given in table~\ref{tab:Tavg-Tsurface}.

\begin{table}[!tbph]
\centering{}\caption{Sample surface temperatures \sym{T}{surface}, average sample temperatures
\sym{T}{avg}, statistic and systematic uncertainty derived from the
FEM model of the CORC sample.\label{tab:Tavg-Tsurface}}
{\scriptsize{}}%
\begin{tabular}{cccc}
\toprule 
{\scriptsize{}surface temperature, \sym{T}{surface}} & {\scriptsize{}average cable temperature, \sym{T}{average}} & {\scriptsize{}statistic uncertainty} & {\scriptsize{}systematic uncertainty}\tabularnewline
\midrule
\midrule 
{\scriptsize{}{\scriptsize{}\SI{15.2}{\kelvin}}} & {\scriptsize{}{\scriptsize{}\SI{14.8}{\kelvin}}} & {\scriptsize{}{\scriptsize{}\SI{\pm 0.3}{\kelvin}}} & {\scriptsize{}{\scriptsize{}\SI{\pm 0.5}{\kelvin}}}\tabularnewline
\midrule 
{\scriptsize{}{\scriptsize{}\SI{21.4}{\kelvin}}} & {\scriptsize{}{\scriptsize{}\SI{20.8}{\kelvin}}} & {\scriptsize{}{\scriptsize{}\SI{\pm 0.5}{\kelvin}}} & {\scriptsize{}{\scriptsize{}\SI{\pm 0.7}{\kelvin}}}\tabularnewline
\midrule 
{\scriptsize{}{\scriptsize{}\SI{33.5}{\kelvin}}} & {\scriptsize{}{\scriptsize{}\SI{32.7}{\kelvin}}} & {\scriptsize{}{\scriptsize{}\SI{\pm 0.7}{\kelvin}}} & {\scriptsize{}{\scriptsize{}\SI{\pm 1.1}{\kelvin}}}\tabularnewline
\midrule 
{\scriptsize{}{\scriptsize{}\SI{42.1}{\kelvin}}} & {\scriptsize{}{\scriptsize{}\SI{41.2}{\kelvin}}} & {\scriptsize{}{\scriptsize{}\SI{\pm 0.9}{\kelvin}}} & {\scriptsize{}{\scriptsize{}\SI{\pm 1.4}{\kelvin}}}\tabularnewline
\midrule 
{\scriptsize{}{\scriptsize{}\SI{44.6}{\kelvin}}} & {\scriptsize{}{\scriptsize{}\SI{43.7}{\kelvin}}} & {\scriptsize{}{\scriptsize{}\SI{\pm 0.8}{\kelvin}}} & {\scriptsize{}{\scriptsize{}\SI{\pm 1.5}{\kelvin}}}\tabularnewline
\midrule 
{\scriptsize{}{\scriptsize{}\SI{49.9}{\kelvin}}} & {\scriptsize{}{\scriptsize{}\SI{59.1}{\kelvin}}} & {\scriptsize{}{\scriptsize{}\SI{\pm 0.8}{\kelvin}}} & {\scriptsize{}{\scriptsize{}\SI{\pm 1.6}{\kelvin}}}\tabularnewline
\midrule 
{\scriptsize{}{\scriptsize{}\SI{55.9}{\kelvin}}} & {\scriptsize{}{\scriptsize{}\SI{55.0}{\kelvin}}} & {\scriptsize{}{\scriptsize{}\SI{\pm 0.8}{\kelvin}}} & {\scriptsize{}{\scriptsize{}\SI{\pm 1.8}{\kelvin}}}\tabularnewline
\midrule 
{\scriptsize{}{\scriptsize{}\SI{62.4}{\kelvin}}} & {\scriptsize{}{\scriptsize{}\SI{61.5}{\kelvin}}} & {\scriptsize{}{\scriptsize{}\SI{\pm 0.8}{\kelvin}}} & {\scriptsize{}{\scriptsize{}\SI{\pm 1.9}{\kelvin}}}\tabularnewline
\midrule 
{\scriptsize{}{\scriptsize{}\SI{69.7}{\kelvin}}} & {\scriptsize{}{\scriptsize{}\SI{68.8}{\kelvin}}} & {\scriptsize{}{\scriptsize{}\SI{\pm 0.6}{\kelvin}}} & {\scriptsize{}{\scriptsize{}\SI{\pm 2.1}{\kelvin}}}\tabularnewline
\midrule 
{\scriptsize{}{\scriptsize{}\SI{73.7}{\kelvin}}} & {\scriptsize{}{\scriptsize{}\SI{72.7}{\kelvin}}} & {\scriptsize{}{\scriptsize{}\SI{\pm 0.7}{\kelvin}}} & {\scriptsize{}{\scriptsize{}\SI{\pm 2.1}{\kelvin}}}\tabularnewline
\midrule 
{\scriptsize{}{\scriptsize{}\SI{75.7}{\kelvin}}} & {\scriptsize{}{\scriptsize{}\SI{74.7}{\kelvin}}} & {\scriptsize{}{\scriptsize{}\SI{\pm 0.6}{\kelvin}}} & {\scriptsize{}{\scriptsize{}\SI{\pm 2.1}{\kelvin}}}\tabularnewline
\midrule 
{\scriptsize{}{\scriptsize{}\SI{77.8}{\kelvin}}} & {\scriptsize{}{\scriptsize{}\SI{76.9}{\kelvin}}} & {\scriptsize{}{\scriptsize{}\SI{\pm 0.5}{\kelvin}}} & {\scriptsize{}{\scriptsize{}\SI{\pm 2.2}{\kelvin}}}\tabularnewline
\bottomrule
\end{tabular}
\end{table}

\begin{figure}[tbh]
\begin{centering}
\includegraphics[width=0.5\textwidth]{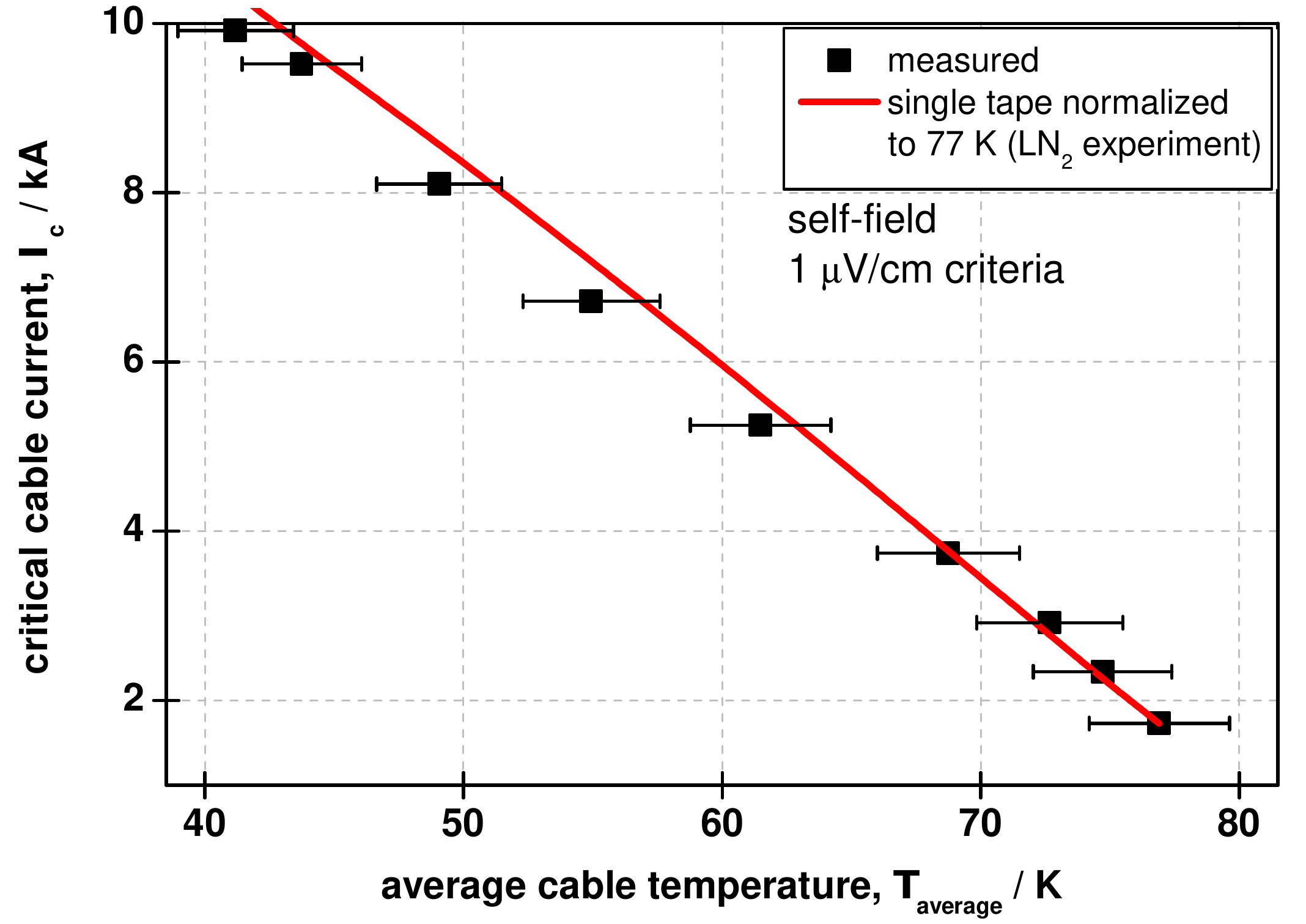}
\par\end{centering}

\caption{Temperature dependence of the critical current of the CORC cable sample
(black dots) and extrapolated \YBCO tapes from SuperPower (red line)
at self-field conditions. The CORC cable is measured with the variable
temperature insert of the FBI test facility. Its average temperature
is calculated using the simulated temperature distribution presented
in figure~\ref{fig:CORC-T-distribution}. Single tape temperature
dependence from~\cite{Hazelton2012-Napa-Workshop}, normalized to
the cables \SI{77}{\kelvin} current carrying capabilities (LN\textsubscript{2}
experiment, see subsection~\ref{sec:77K-characterization}).\label{fig:CORC-self-field-T-dependence}}
\end{figure}

The measured cable critical current \sym{I}{c} depending on the average
temperature \sym{T}{average} is compared in figure~\ref{fig:CORC-self-field-T-dependence}
with the single \YBCO tape temperature dependence published by the
manufacturer \cite{Hazelton2012-Napa-Workshop}. Single tapes are
extrapolated by normalizing to the CORC cable's \SI{77}{\kelvin}
current carrying capabilities (from the LN\textsubscript{2} bath,
self-field experiment in subsection~\ref{sec:77K-characterization}).
Increased self-fields at lower temperatures and higher cable currents
are not considered in this extrapolation.

The temperature variable CORC cable measurement and the extrapolated
single tape data are in good agreement within the uncertainty of the
temperature determination (combining statistic and systematic uncertainties,
see table~\ref{tab:Tavg-Tsurface}). In the investigated temperature
range of \SIrange{40}{77}{\kelvin}, the temperature dependence is
linear for the single tapes and nearly linear for the CORC cable measurement.

\subsection{Magnetic field cycles\label{sec:Load-cycling}}

In the third step, the CORC cable sample was checked for degradation
of the current carrying capabilities due to high transverse mechanical
loads. This was achieved through cycling of the magnetic background
field, currents and therefore transverse loads generated by the Lorentz
force. The magnetic background field was increased from \SI{2}{\tesla}
(due to limitations of the current source, maximum current is \SI{10}{\kilo\ampere})
to \SI{12}{\tesla}. Every \SI{2}{\tesla}, the critical current was
measured at \SI{4.2}{\kelvin} using a \SI{1}{\micro\volt\per\centi\metre}
criteria with the voltage tap pair at the copper contacts and assuming
\SI{70}{\milli\metre} (magnet's high field region) to calculate electric
fields while subtracting any linear ohmic contribution.

\begin{figure}[tbh]
\begin{centering}
\includegraphics[width=0.5\textwidth]{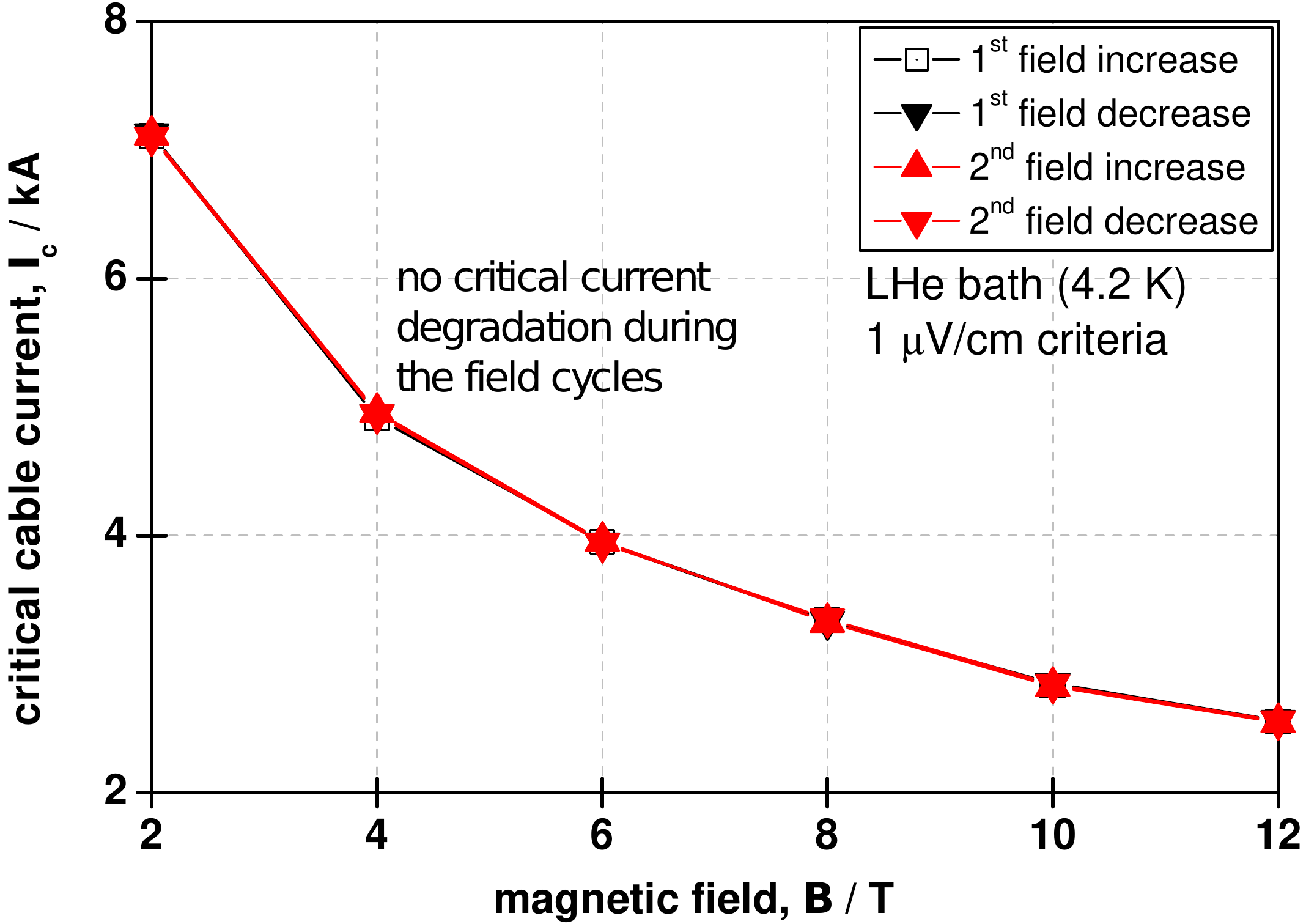}
\par\end{centering}

\caption{Critical current measurements at increasing and decreasing magnetic
background fields of the CORC cable sample. There is no degradation
of the sample's current carrying capabilities in different magnetic
field cycles.\label{fig:Load-cycling}}
\end{figure}

The critical current index values (n-values) of all superconducting
transitions are in the \SIrange{39}{55}{} range. N-values are obtained
between the critical electric field (\SI{1}{\micro\volt\per\centi\metre})
and ten times the critical electric field (\SI{10}{\micro\volt\per\centi\metre})
using nonlinear fitting and assuming power law behavior. The current
carrying capabilities are identical on the first (black curves in
figure~\ref{fig:Load-cycling}) to the following field cycle (red
curves). During the field cycles, the maximal transverse Lorentz forces
(per sample length) on the CORC cable sample of \SI{31.4}{\kilo\newton\per\metre}
occur during the critical current measurements at \SI{12}{\tesla}
magnetic background field. No irreversible degradation of the current
carrying capabilities is observed, the sample's performance remains
constant during all field cycles. 

\subsection{Magnetic field- and temperature dependence\label{sec:T-variable+field-variable}}

In the fourth step, the magnetic field- and temperature dependence
of the current carrying capabilities of the CORC cable sample were
determined. The sample's critical currents were measured for sample
surface temperatures from \SI{4.2}{\kelvin} to \SI{51.5}{\kelvin}
in the magnetic field range from \SI{2}{\tesla} to \SI{12}{\tesla}.
The critical currents were measured each \SI{2}{\tesla} for different
cable surface temperatures with a \SI{1}{\micro\volt\per\centi\metre}
criteria at the copper contacts' voltage tap pair. A distance of \SI{70}{\milli\metre}
(high temperature \& high field region) was utilized to calculate
electric fields while subtracting the linear ohmic contribution of
the contact resistance. The same sample surface temperature steps
were used in the whole magnetic field range.

\subsubsection{Behavior of the sample\label{sub:Sample-behaviour}}

\begin{figure*}[tp]
\begin{centering}
\includegraphics[width=0.5\textwidth]{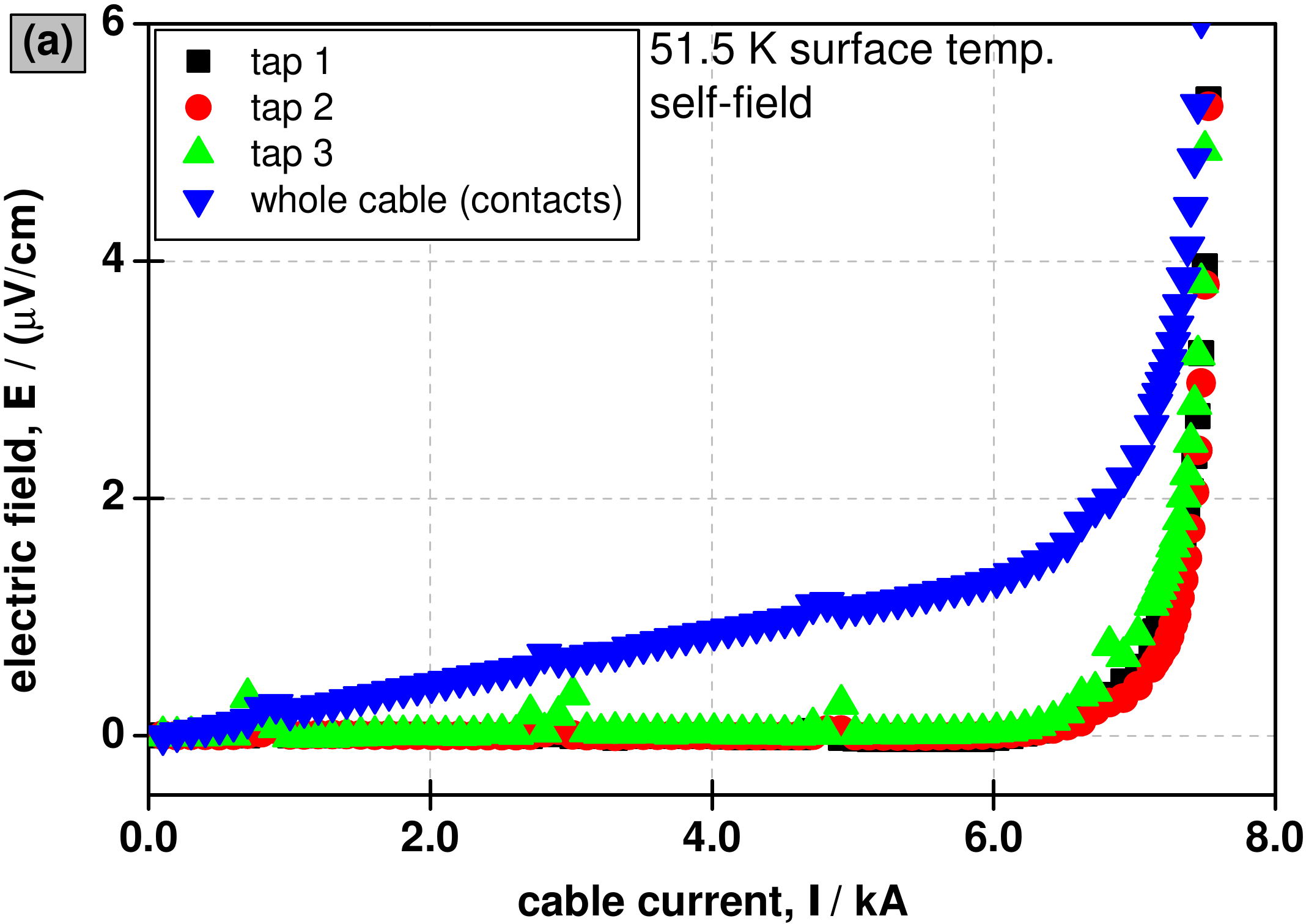}\includegraphics[width=0.5\textwidth]{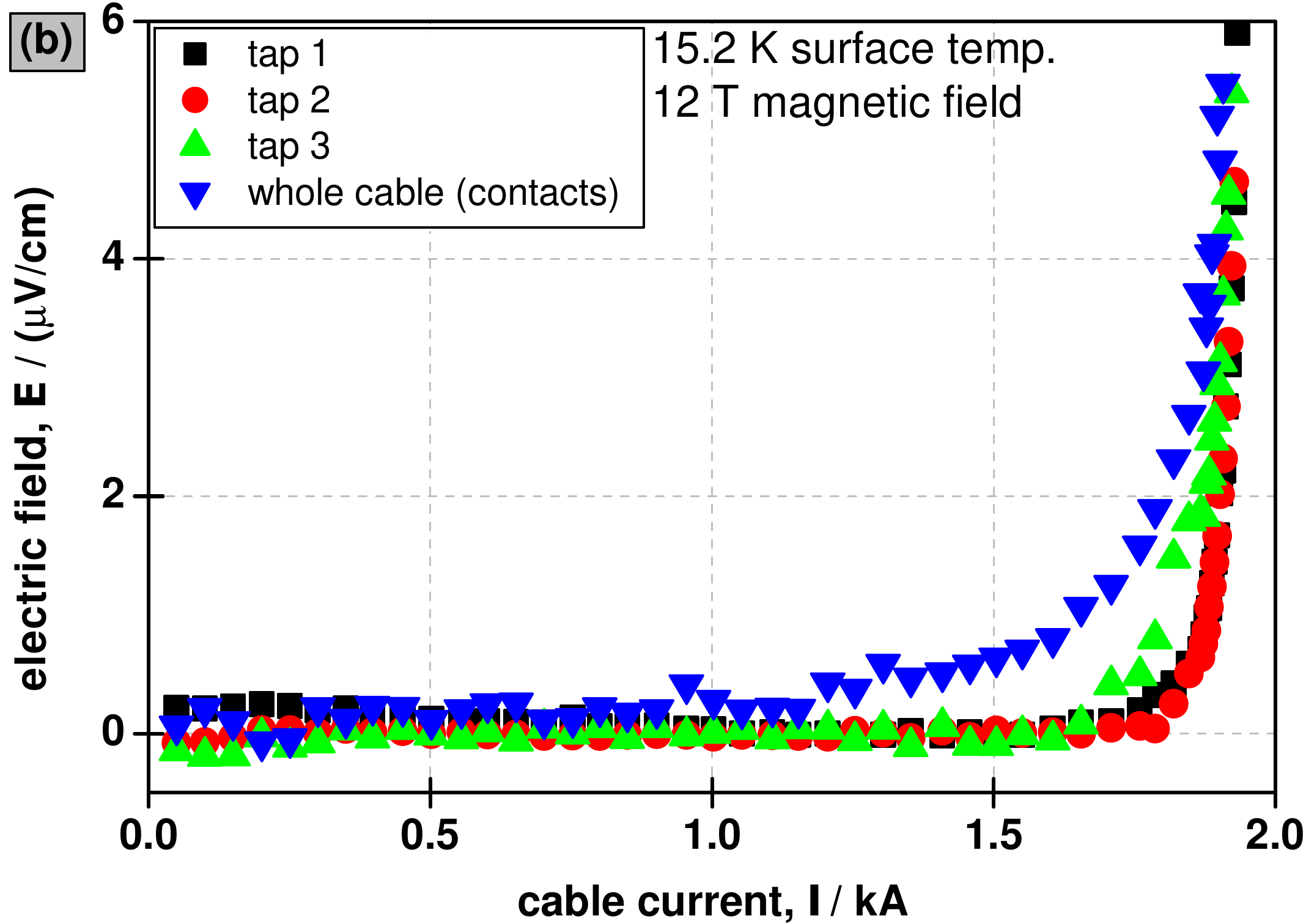}
\par\end{centering}

\caption{Exemplary electric field vs. current curves for the superconducting
transitions in the field- and temperature dependent measurement of
the CORC cable sample: high surface temperature (left: a) and high
magnetic background field (right: b).\label{fig:CORC-in-field-E-vs-I}}
\end{figure*}

Regardless of magnetic field or temperature, the shape of the transitions
from superconducting state to normal conduction remain similar; all
transitions are steep. In figure~\ref{fig:CORC-in-field-E-vs-I},
two exemplary electric field vs. current curves are shown: high temperature
(left:~a) and high magnetic background field (right:~b). All data
is as measured, a distance of \SI{70}{\milli\metre} is used to calculate
electric fields and linear contributions are not subtracted. Voltage
taps on the \YBCO tape exhibit minimal linear ohmic contributions,
while at the copper contact these are due to the contact resistances
more pronounced. The overall contact resistance is \SI{1.6}{\nano\ohm},
it is obtained by comparing the linear part's slope of the electric
field vs. cable current curves at the voltage taps on the \YBCO tapes
with the behavior of the whole cable, measured at the copper terminations.
Using the previously described calculation method (see subsection~\ref{sec:T-variable}),
the current bypass is negligible and remains below \SI{4}{\ampere}
even at ten times the critical electric field.

Similar behavior was obtained for all superconducting transitions
of this sample, implying constant performance in the whole magnetic
field- and temperature range. The sample's electrical stabilization
is sufficient as no significant increase of the cable temperature
is observed during superconducting transitions even close to the maximal
current of the test facility (\SI{10}{\kilo\ampere}).

\subsubsection{Critical current vs. magnetic field and surface temperature}

\begin{figure}[tbh]
\begin{centering}
\includegraphics[width=0.5\textwidth]{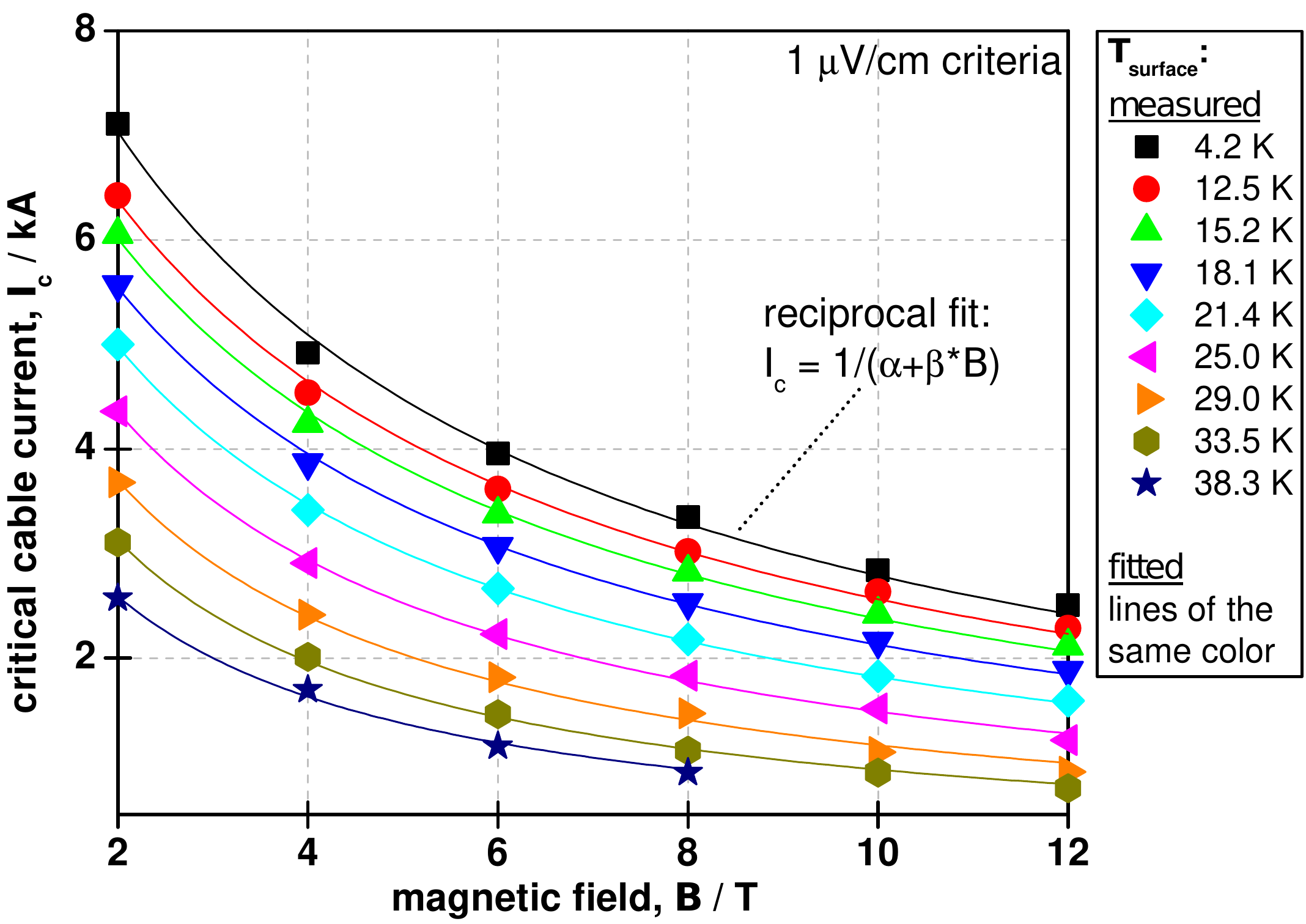}
\par\end{centering}

\caption{Magnetic field- and surface temperature \sym{T}{surface} dependent
measurements of the CORC cable sample. The measured data (points)
is fitted with reciprocal functions (lines).\label{fig:CORC-surface_temp-and-field}}
\end{figure}

The obtained critical cable currents (y-axis) at different magnetic
background fields (x-axis) and sample surface temperatures (in the
legend) are shown in figure~\ref{fig:CORC-surface_temp-and-field}.
The obtained field dependent curves of different sample surface temperatures
are regularly spaced and follow a reciprocal dependence (\ensuremath{\sym{I}{c} = \frac{1}{\left(\alpha+\beta\cdot B\right)}}). 

\subsubsection{Critical current vs. magnetic field and average cable temperature}

\begin{figure}[tbh]
\begin{centering}
\includegraphics[width=0.5\textwidth]{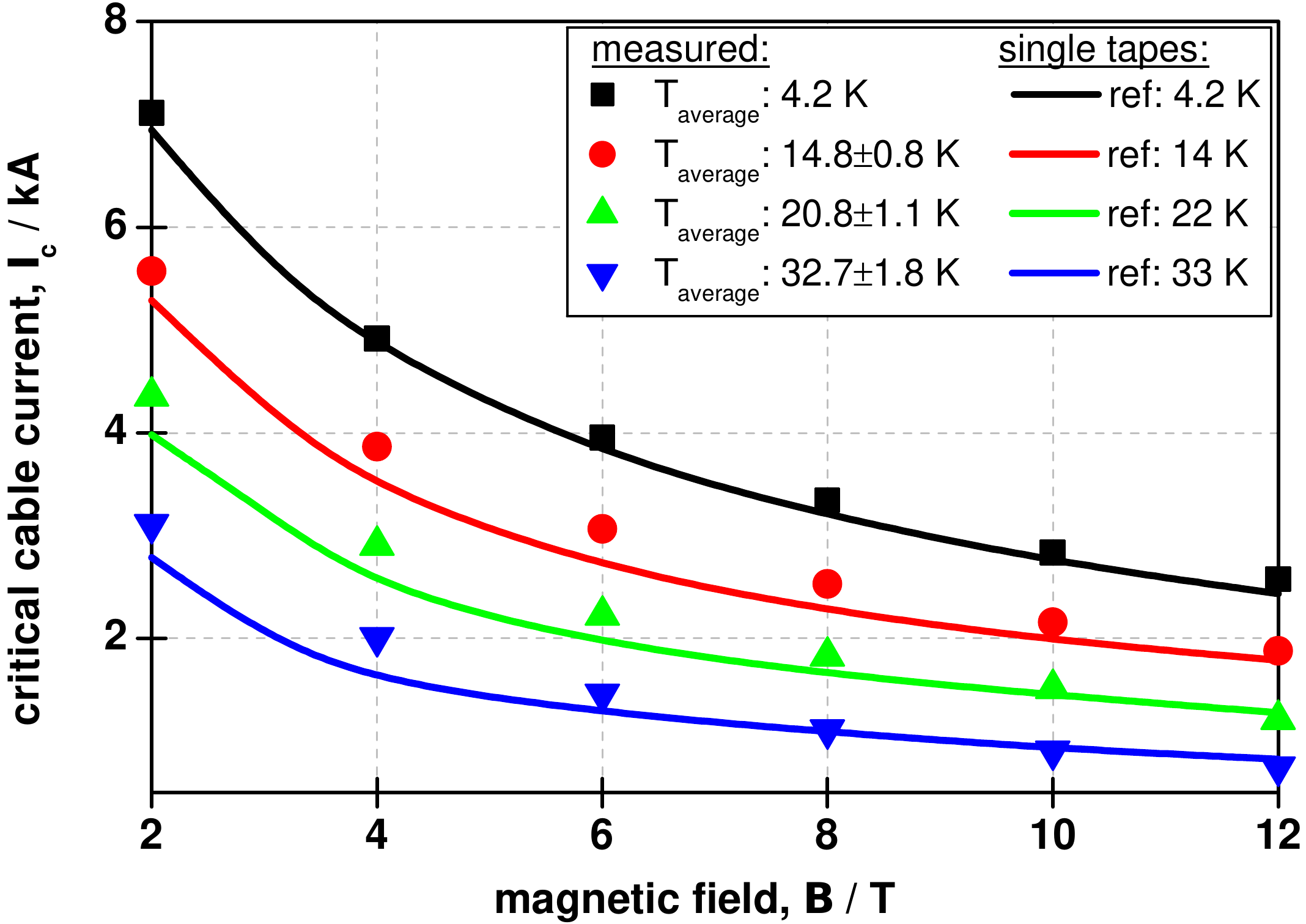}
\par\end{centering}

\caption{Field- and average cable temperature \sym{T}{average} dependent measurements
of the CORC cable sample: measurements (points) and extrapolated single
tape data (lines). The \SI{4.2}{\kelvin}, \SI{12}{\tesla} current
carrying capabilities of \YBCO tapes from the same batch are extrapolated
to different fields and temperatures using the single tape field-
and temperature dependence published by the manufacturer \cite{Hazelton2012-Napa-Workshop}.\label{fig:CORC-eff_temp-and-field}}
\end{figure}

Using the simulated temperature dependence of the CORC cable sample
(subsection~\ref{sec:T-distribution}) and averaging the temperature
of the tapes for each position, an average cable temperature was calculated
for the different sample surface temperatures. Surface temperatures
\sym{T}{surface}, average temperatures \sym{T}{average} as well
as the statistic (derived from the averaging) and the systematic (from
the temperature measurement) uncertainties are given in table~\ref{tab:Tavg-Tsurface}.
The critical currents at different magnetic background fields (x-axis)
and effective temperatures (in the legend) are compared with the extrapolated
current carrying capabilities of single \YBCO tapes. The average
\SI{4.2}{\kelvin}, \SI{12}{\tesla} current carrying capabilities
of the sample's \YBCO tapes (\SI{161.42}{\ampere} measured on tapes
of the same batch, see table~\ref{tab:Sample-parameters}) are extrapolated
to different fields and temperatures using the single tape field-
and temperature dependence factors \ensuremath{I_{\n{c}}(T,B_{\perp})/I_{\n{c}}(\SI{4.2}{\kelvin},\SI{12}{\tesla}_{\perp})}
published by the \YBCO tapes' manufacturer \cite{Hazelton2012-Napa-Workshop}.
The magnetic background field as well as the cable's self-field are
considered. Measured data (points) and extrapolated data (lines) are
shown in figure~\ref{fig:CORC-eff_temp-and-field}.

Measured and extrapolated data are in good agreement, with the CORC
cable currents being slightly higher than the extrapolated \YBCO
tapes in the majority of field and temperature combinations. At \SI{4.2}{\kelvin},
and \SI{12}{\tesla}\ensuremath{\;\perp} (perpendicular magnetic background
field) the CORC cable sample carries \SI{2.53}{\kilo\ampere} corresponding
to a linear (one dimensional) current density of \SI{422}{\ampere\per\centi\metre-width}.
This perfectly matches the performance reported by the manufacturer
reported for this type of conductor (SCS4050):\SI{425}{\ampere\per\centi\metre-width}
at \SI{4.2}{\kelvin}, \SI{12}{\tesla}\ensuremath{\;\perp} \cite[p. 8]{Hazelton2012-Napa-Workshop}\cite{SuperPower-2G-wire}).
This clearly shows that the \YBCO tapes' current carrying capabilities
are fully utilized in the CORC cable concept and that the sample has
a temperature- and magnetic background field dependence identical
to single tapes.

\subsection{High current ramp rates\label{sub:CORC-high-ramp-rates}}

\begin{figure}[tbh]
\begin{centering}
\includegraphics[width=0.5\textwidth]{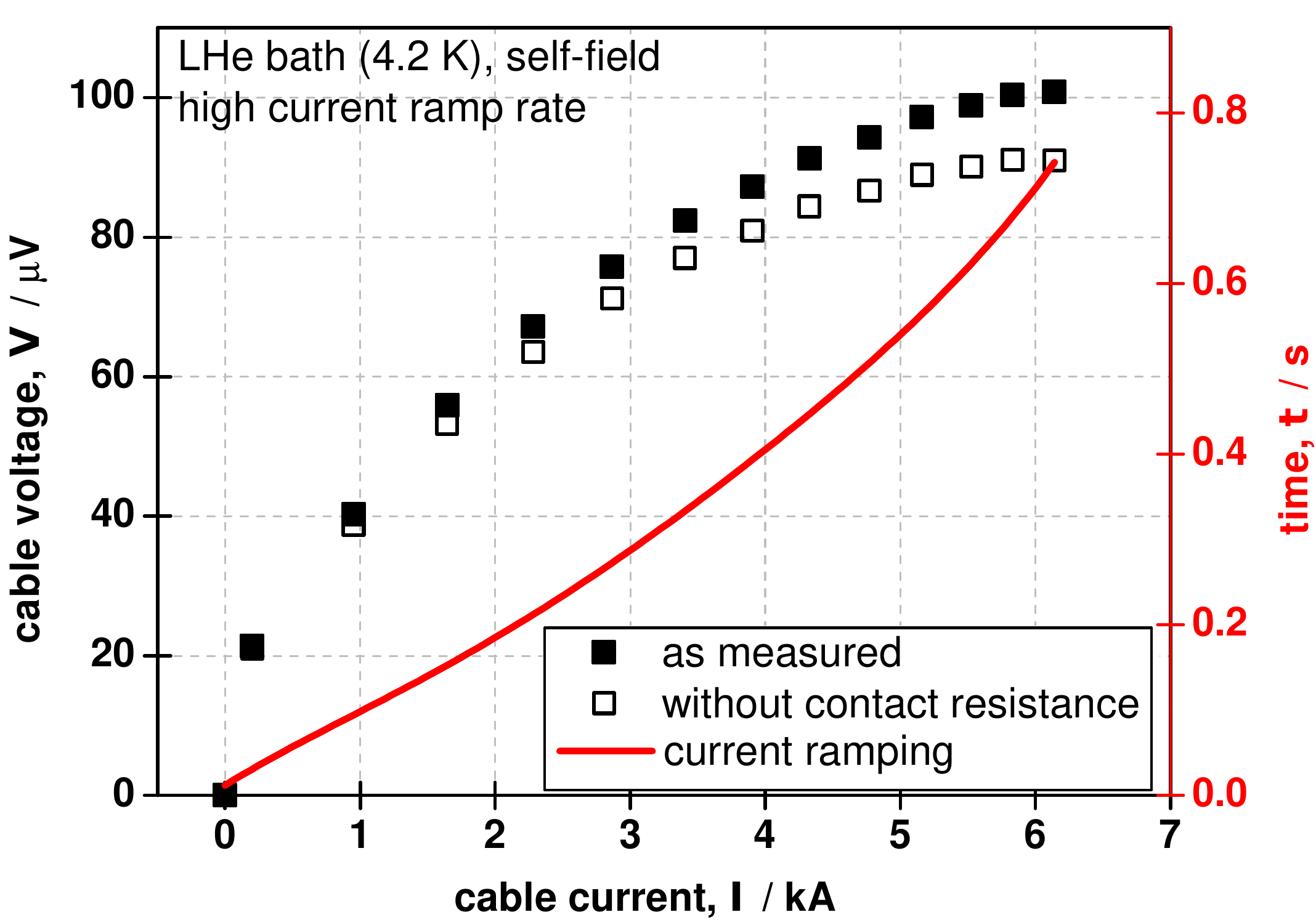}
\par\end{centering}

\caption{High current ramp rates on CORC cables. Current of the sample (\SI{4.2}{\kelvin},
self-field conditions) is increased at a ramp rate of \SI{8.3}{\kilo\ampere\per\second}.
The voltage drop between the copper terminations is measured with
a fast data acquisition system. The measured voltages (full symbols)
and the voltage without the contribution of the contact resistance
(hollow symbols) as well as the current ramping (line) are shown.\label{fig:CORC-high-current-ramp-rates}}
\end{figure}

In the fifth step, the response of the CORC cable sample to rapidly
changing currents was investigated to simulate the fast charging or
discharging of magnets. Using a low noise signal amplifier%
\footnote{voltage amplifier with a \SI{3}{\deci\bel} frequency of \SI{300}{\hertz}%
} and a data acquisition system%
\footnote{data acquisition rate of \SI{100}{\hertz}%
}, the voltage across the CORC cable sample's copper contacts was measured
during the fast ramping of the current (current ramp rate of \SI{8.3}{\kilo\ampere\per\second}).
The measurement was done at self-field conditions and a cable temperature
of \SI{4.2}{\kelvin}. At the boundary conditions, the sample's critical
current was above the test facilities maximum current, the measured
voltage was therefore the result of the contact resistance (\SI{1.6}{\nano\ohm},
see subsection~\ref{sub:Sample-behaviour}) as well as the sample's
impedance. The measured voltages (full symbols) and the voltage without
the contribution of the contact resistance (hollow symbols) as well
as the current ramping (line) are shown in figure~\ref{fig:CORC-high-current-ramp-rates}.

No voltage spikes are observed. The maximal occurring voltage over
the whole sample (length of \SI{0.96}{\metre} between the terminations,
see table~\ref{tab:Sample-parameters}) is \SI{1.008E-4}{\volt}
and \SI{0.910E-4}{\volt} subtracting the contribution of the contact
resistance. Combined with the average current ramp rate of \SI{8.3}{\kilo\ampere\per\second},
the sample's inductance is of the order of \SI{1E-8}{\henry}. The
corresponding electric field even remains below the critical electric
field of \SI{1}{\micro\volt\per\centi\metre}. At this voltage the
current bypass through the copper former is still negligible, it remains
below \SI{5.3}{\ampere} using the previously described calculation
method (see subsection~\ref{sec:T-variable}). The investigated CORC
cable sample is able to withstand fast changes of its current.

\section{Discussion and conclusion}

The temperature and magnetic field depended current carrying capabilities
of a Conductor on Round Core (CORC) cable with 15 SuperPower SCS4050
(no advanced pinning) tapes (3 per layer) have been investigated in
detail using the FBI test facility of the Institute for Technical
Physics (ITEP) of the Karlsruhe Institute of Technology (KIT). A variable
temperature insert locally heated the CORC cable sample. This system's
temperature distribution was modeled in fully scale with FEM using
the direction and temperature dependent thermal conductivities of
the sample's constituent materials. Constant and maximal temperatures
were found in a \SI{70}{\milli\metre} long region, the high temperature
region corresponding to the high temperature region of the magnet.
Outside this region, temperatures and magnetic fields dropped quickly
and \SI{70}{\milli\metre} were therefore used to convert voltages
into electric fields in all temperature- and magnetic field dependent
measurements. The CORC cable sample's superconducting transitions
were steep and of high n-value (\SIrange{39}{55}{} range) allowing
the use of a \SI{1}{\micro\volt\per\centi\metre} criteria. The current
bypass through the copper in the former is negligible. The current
carrying capabilities of the \YBCO tapes from the same batch have
been determined (on several single tapes, not assembled to a cable).
In average, each of the \SI{4}{\milli\metre} wide tapes carried \SI{127.57}{\ampere}
at \SI{77}{\kelvin}, self-field and \SI{161.42}{\ampere} at \SI{4.2}{\kelvin}
\SI{12}{\tesla}.

The investigations were performed in five steps. In the first step
the sample's \SI{77}{\kelvin} (LN\textsubscript{2} bath), self-field
current was determined to \SIrange{1652}{1757}{\ampere} depending
on the used voltage tap. Due to these differences, all following experiments
use the voltage taps on the copper contacts (the sample's terminals)
the averages over the behavior of all tapes. Referring to this voltage
tap, the CORC cable carries \SI{1638}{\ampere}; compared with its
design value of \SI{1913}{\ampere}, this corresponded to a self-field
degradation of only \SI{14.38}{\percent}. In the second step, the
sample's self-field temperature dependence was determined in the \SIrange{42}{77}{\kelvin}
range and compared with extrapolated single tapes. The CORC cable's
temperature is derived by combining the simulated temperature distribution
with the measured sample surface temperature. The single tape extrapolation
uses the tapes' temperature dependence as published by the manufacturer
and normalizes with the \SI{77}{\kelvin} (LN\textsubscript{2} bath),
self-field cable current. Measured and extrapolated data are in good
agreement within the uncertainty of the determination of the average
cable temperature. In the third step, the sample's stability against
high transverse mechanical loads was investigated by repeatedly cycling
the magnetic background field between \SI{2}{\tesla} and \SI{12}{\tesla}.
No degradation of its current carrying capability was observed during
the magnetic field cycles with maximal transverse Lorentz forces (per
sample length) of up to \SI{31.4}{\kilo\newton\per\metre}. In the
fourth step, the magnetic field- and temperature dependence of the
current carrying capabilities of the CORC cable sample were determined
and compared with extrapolated single tape data. The extrapolation
uses the average \SI{4.2}{\kelvin}, \SI{12}{\tesla} current carrying
capabilities\YBCO tapes from the same batch and the field- and temperature
dependence factors \ensuremath{I_{\n{c}}(T,B_{\perp})/I_{\n{c}}(\SI{4.2}{\kelvin},\SI{12}{\tesla}_{\perp})}
published by the tapes' manufacturer. Measured and extrapolated data
are in good agreement, with the CORC cable's currents being slightly
higher than the extrapolated \YBCO tapes in the majority of field
and temperature combinations. At \SI{4.2}{\kelvin}, and \SI{12}{\tesla}\ensuremath{\;\perp}
(perpendicular magnetic background field) the sample has a critical
current of \SI{2.53}{\kilo\ampere} corresponding to a current density
of \SI{422}{\ampere\per\centi\metre-width} which perfectly matches
the performance reported by the manufacturer. This clearly shows that
the sample's temperature- and field dependence is identical to that
of its constituent \YBCO tapes as well as that tapes' current carrying
capabilities are fully utilized. In the fifth step, the CORC cable's
response to rapidly changing currents was investigated. No voltage
spikes were observed at current ramp rates of \SI{8.3}{\kilo\ampere\per\second},
the inductance is of the order of \SI{1E-8}{\henry}. Subtracting
the contact resistance of the copper terminations (\SI{1.6}{\nano\ohm}),
the electric field remains below the critical electric field of \SI{1}{\micro\volt\per\centi\metre}.
There was no irreversible reduction of the CORC cable's performance
in any of the experimental steps.

These finding clearly show CORC cables being very promising candidates
as conductor in high field or fusion magnets.

\ack{}{}

This work is supported in part by the U.S. Department of Energy, under
contract numbers DE-AI05-98OR22652 and DE-SC0007660. The authors acknowledge
Frank Gr\"{o}ner, Valentin Tschan, Sascha Westenfelder and Anton
Lingor. Only their technical support made these experiments possible.

\bibliographystyle{unsrt}
\addcontentsline{toc}{section}{\refname}\bibliography{library}

\end{document}